\documentclass[lettersize,journal]{IEEEtran}
\usepackage{amsmath,amsfonts}
\usepackage{algorithmic}
\usepackage{algorithm}
\usepackage{array}
\usepackage[font=small,labelfont=bf,labelsep=colon,justification=centering]{caption}
\usepackage{textcomp}
\usepackage{stfloats}
\usepackage{url}
\usepackage{verbatim}
\usepackage{graphicx}
\usepackage{cite}

\usepackage{siunitx}
\usepackage{amsmath,amssymb,amsfonts}
\usepackage{algorithmic}
\usepackage{graphicx}
\usepackage{textcomp}
\usepackage{hyperref}
\usepackage{enumitem}
\usepackage{amssymb}
\usepackage{ragged2e}
\usepackage{algorithm}
\usepackage{multicol}
\usepackage{soul}
\usepackage{array} 
\usepackage{booktabs} 
\usepackage{listings}
\usepackage{array}
\usepackage{multirow}
\usepackage{times}
\usepackage{stfloats}
\usepackage{colortbl}
\usepackage{amsmath}
\newcolumntype{C}[1]{>{\centering\arraybackslash}m{#1}}

\usepackage{booktabs}
\usepackage{tabularx}
\usepackage{arydshln}
\usepackage[table]{xcolor}
\arrayrulecolor{gray}

\begin{document}

\title{F-RBA: A Federated Learning-based Framework for Risk-based
Authentication}

\author{
Hamidreza Fereidouni, Abdelhakim Senhaji Hafid, Dimitrios Makrakis, Yaser Baseri
\thanks{Hamidreza Fereidouni, Abdelhakim Senhaji Hafid, and Yaser Baseri are with the  Department of Computer Science and Operations Research, University of Montreal, Quebec, Canada.}
\thanks{Dimitrios Makrakis is with the School of Electrical Engineering and Computer Science, University of Ottawa, Ontario, Canada.}
\thanks{Correspondence: H. Fereidouni (\href{mailto:hamidreza.fereidouni@umontreal.ca}{hamidreza.fereidouni@umontreal.ca})}
\thanks{This work has been supported by the \href{https://www.nserc-crsng.gc.ca/}{Natural Sciences and Engineering Research Council of Canada (NSERC)} and \href{https://flexgroups.com}{Flex Group}.}
}


\markboth{}%
{Shell \MakeLowercase{\textit{et al.}}: A Sample Article Using IEEEtran.cls for IEEE Journals}

\IEEEpubid{}

\maketitle

\begin{abstract}
The proliferation of Internet services has led to an increasing need to protect private data. User authentication serves as a crucial mechanism to ensure data security. Although robust authentication forms the cornerstone of remote service security, it can still leave users vulnerable to credential disclosure, device-theft attacks, session hijacking, and inadequate adaptive security measures. Risk-based Authentication (RBA) emerges as a potential solution, offering a multi-level authentication approach that enhances user experience without compromising security. In this paper, we propose a Federated Risk-based Authentication (F-RBA) framework that leverages Federated Learning to ensure privacy-centric training, keeping user data local while distributing learning across devices. Whereas traditional approaches rely on centralized storage, F-RBA introduces a distributed architecture where risk assessment occurs locally on users' devices. The core innovation of the framework lies in its similarity-based feature engineering approach, which addresses the heterogeneous data challenges inherent in federated settings—a significant advancement for distributed authentication. By facilitating real-time risk evaluation across devices while maintaining unified user profiles, F-RBA achieves a balance between data protection, security, and scalability. Through its federated approach, F-RBA addresses the cold-start challenge in risk model creation, enabling swift adaptation to new users without compromising security. Empirical evaluation using a real-world multi-user dataset demonstrates the framework's effectiveness, achieving a superior true positive rate for detecting suspicious logins compared to conventional unsupervised anomaly detection models. This research introduces a new paradigm for privacy-focused RBA in distributed digital environments, facilitating advancements in federated security systems.
\end{abstract}



\section{Introduction}

In the age of the Internet, it is crucial to protect against identity compromise. Authentication is the process of confirming that users, processes, or devices are as legitimate as they claim to be, involving identity verification before allowing access to an information system's resources \cite{nist_authentication}. The verification is performed using one or a combination of authentication techniques, including (a) something the user knows, such as passwords; (b) something the user possesses, like security tokens; and (c) something the user is, such as biometric identifiers \cite{hazratifard2022using}. Advanced authentication systems use context-aware verification \cite{benzekki2018context}, considering factors like IP address, geolocation, time of access, and device information. This method allows for seamless authentication without requiring the user's attention by monitoring user behavior and contextual factors \cite{oza2019active}. Securing reliable authentication systems becomes increasingly challenging, as digital footprints expand for individuals and organizations alike. Strong authentication protocols are essential, serving as the primary defense against unauthorized access.

The cyber threat landscape evolves rapidly, outpacing the advancements in authentication technology. The IBM X-Force Threat Intelligence Index 2024 \cite{ibm_xforce_2024} reported a 71\% increase in cyberattacks utilizing stolen credentials over the previous year. A 2022 study by \cite{gavazzi2023study}, analyzing 208 sites from the Tranco top 5K, revealed that only 42.31\% implemented Multi-Factor Authentication (MFA), while merely 22.12\% blocked suspicious login attempts, highlighting significant vulnerabilities in current practices. Market trends reflect growing demand for advanced authentication solutions. The global Risk-based Authentication market is projected to grow from \$5.0 billion USD in 2023 to \$16.5 billion USD by 2032, representing a Compound Annual Growth Rate (CAGR) of 14.2\% \cite{risk_based_authentication_market}. This data underscores the urgent need for implementing adaptive security measures to address current vulnerabilities and emerging threats in authentication systems.

Risk-based Authentication (RBA) is an adaptive and dynamic user authentication process that assesses the risk level associated with each access attempt in real-time. It evaluates various contextual and behavioral factors, such as device type, user location, time of access, and user's behavioral patterns. Based on these parameters, RBA adjusts its defenses and verification requirements accordingly; it allows the system to protect against potential threats, apply appropriate authentication techniques, and enhance security while maintaining user experience \cite{wiefling2019really}. By responsively opting for appropriate security measures, RBA strikes a balance between robust security and user convenience, adapting to the specific risk profile of each authentication attempt \cite{wiefling2020more}. For instance, accessing sensitive data from an unknown device might require additional verification. Despite the fact that RBA is superior to traditional static authentication methods, challenges remain, particularly in balancing scalability, user data privacy, and performance. Numerous strategies can strengthen RBA, including but not limited to using multiple ML models for accurate risk determination \cite{buriro2021risk, liu2022log}, deploying on-device risk engine \cite{liu2022log, picard2023rlauth}, and employing specialized approaches for certain applications, such as using RBA for border control machines \cite{papaioannou2022toward}.

Current RBA systems face significant limitations. Most are server-centric, requiring transmission of raw contextual and behavioral data to authentication servers. Some systems do not employ AI-driven models for risk assessment, limiting automated pattern recognition. Several systems fail to support continuous learning from user context and behavior, hindering model adaptation to new patterns. Additionally, these frameworks often face cold start challenges due to insufficient historical data, resulting in suboptimal security decisions for new users or contexts. To the best of our knowledge, there is no contribution in the open literature that reports the combination of a distributed risk engine for model training—essential for user data privacy preservation—with improved scalability.

\subsection{Contributions of the Work}
\label{subsec:Contributions of the Work}

This paper aims to propose an innovative RBA framework, focusing on the risk engine component to address the above-mentioned limitations. Our contributions can be summarized as follows:
\begin{itemize} [leftmargin=*]
\item \textbf{{Scalable Privacy-Centric Training}}:  We introduce the use of Federated Learning in RBA by keeping user data local, distributing learning across devices, and sending only model updates to the server. This approach ensures privacy while leveraging collective learning for enhanced security.
\item \textbf{Non-IID and Heterogeneous Data Mitigation}: To address the challenges of user heterogeneity and non-IID data distributions, we employ a specific aggregation algorithm and propose similarity-based feature engineering. This approach represents user features in relation to their historical behavior, thereby improving model consistency and convergence in federated settings.
\item \textbf{IPFS-Enabled Cross-Device User Profiling}:
We leverage InterPlanetary File System (IPFS) to support real-time, cross-device user profiling. By synchronizing risk evaluations across multiple devices, this approach enables more comprehensive and continuous monitoring, an unexplored capability in on-device RBA systems. 
\item \textbf{Cold-Start Resolution}: The framework combines global model aggregation with personalized thresholds. This allows quick adaptation to new users, providing reliable risk assessments from initial login attempts.
\end{itemize}

\subsection{Organization}
\label{subsec:Organization}

The structure of this paper is organized as follows: Section \hyperref[sec:Related Work]{II} provides a comprehensive literature review on RBA, encompassing both foundational research and recent advancements. Section \hyperref[sec:System Model and Facets]{III} delineates the system model, potential threats, and critical aspects of RBA. Section \hyperref[sec:Proposed Framework]{IV} introduces our proposed framework and its architectural design, discussing everything, from features and heterogeneous data mitigation to the federated model and risk assessment. Section \hyperref[sec:Implementation and Analysis]{V} details the implementation results and offers an in-depth analysis of the proposed framework. Section \hyperref[sec:Challenges and Future Research Directions]{VI} explores the challenges encountered and outlines future research directions. Finally, Section \hyperref[sec:Conclusion]{VII} concludes our work and summarizes the key findings.


\section{Related Work}
\label{sec:Related Work}

This section provides an overview of related work on RBA systems and Federated Learning, identifying limitations in existing contributions. Subsequent research has expanded on these foundational ideas, incorporating sophisticated techniques for more accurate risk evaluation. These techniques include multiple Machine Learning (ML) models, privacy-preserving methods for user data, and more advanced features. This review highlights the evolution of RBA systems and the diverse methodologies employed in the field.

\subsection{Risk-based Authentication}
\label{subsec:Risk-based Authentication}

In the literature on RBA, several key information sources are employed to assess risk levels. These involve the user's historical data, context of the current authentication attempt, and asset sensitivity. Each of these factors provides valuable insights independently, and when combined, they contribute to a comprehensive risk assessment engine and an effective overall RBA system.

Balancing security, usability, and privacy in authentication systems has been a key focus of recent research. Wiefling et al. \cite{wiefling2020more} compared RBA variants with Two-factor Authentication (2FA) and password-only authentication, finding RBA as secure as 2FA and more usable, while surpassing password-only in security. A follow-up study \cite{wiefling2021privacy} addressed privacy risks in server-centric systems; they recommended truncation and k-anonymity to enhance user privacy, while noting the security-privacy trade-off, especially regarding IP addresses. While these methods improve server-based systems, an on-device approach could further mitigate data disclosure risks and enhance user control. These studies \cite{wiefling2020more, wiefling2021privacy} provide valuable insights for optimizing RBA systems.

RBA systems employ diverse approaches to user risk assessment, considering various features and device requirements. Li et al. \cite{li2020wrist} focused on behavioral biometrics, including mouse movements, keystrokes, and wrist gestures for user authentication. Rivera et al. \cite{rivera2020risk} proposed an RBA system that utilizes network latency profiles, though relying solely on network features limits its authentication accuracy. Xu et al. \cite{xu2020gait} developed an approach that relies on gait analysis through smartwatches, while Wang et al. \cite{wang2019context} presented a framework that considers only gesture and touch patterns specific to touchscreen systems. Although these frameworks demonstrate strong performance, their device-specific nature inherently limits their applicability.

Domain-specific implementations of RBA have demonstrated promising results across various fields. Papaioannou et al. \cite{papaioannou2022toward} investigated its application for mobile identification devices at border controls; they used multiple contextual and behavioral features for risk evaluation. Other domain-specific applications include a system for dynamic authenticator selection in Point-of-Service (PoS) payments \cite{wojtowicz2017technical} and a biometric rule-based driver authentication system for ride-sharing platforms \cite{gupta2019driverauth}. Although these studies showcase practical applications, their domain-specific designs hinder broader adoption.

Various models have been proposed for risk assessment in authentication systems, ranging from rule-based approaches to mathematical models and ML techniques. Sepczuk et al. \cite{sepczuk2018new} developed a mathematically grounded rule-based engine that overcomes the limitations of static restrictions and baselines by dynamically assigning risk levels to specific contexts, thereby enhancing the security and adaptability of account lockout mechanisms. Ghosh et al. \cite{ghosh2019softauthz} proposed a system that uses predefined mathematical functions and statistical techniques to compute confidence scores.
In contrast, Wu et al. \cite{wu2020caiauth} employed multiple models such as Autoencoder, One-Class Support Vector Machine (OC-SVM), and Local Outlier Factor (LOF) for risk evaluation. Solano et al. \cite{solano2019risk} combined session context data with behavioral biometrics using Random Forest Classifiers. These approaches \cite{wu2020caiauth, solano2019risk} support dynamic adjustment by adapting to behavioral and contextual changes. However, except for \cite{wu2020caiauth}, all these methods remain server-centric. Additionally, \cite{ghosh2019softauthz, wu2020caiauth} support continuous RBA to some extent.

\renewcommand{\arraystretch}{1.2}
\begin{table*}[b]
\vspace{-0.1cm}
\caption{Comparison of F-RBA with other recent RBA systems (DA: Dynamic Adjustment, ACF: Automatic Context Finding, CPR: Collective Pattern Recognition, CRA: Continuous Risk Assessment, AC: Asset Criticality, ODS: On-Device Setup)}
\label{tab:Main-Comparison}
\centering
\small
\resizebox{\linewidth}{!}{ 
\begin{tabular}{p{5.2cm}p{4.8cm}p{4.2cm}C{1.3cm}C{1.3cm}C{1.3cm}C{1.3cm}C{1.3cm}C{1.3cm}}
\hline
\textbf{Related Work} & \textbf{Data Type} & \textbf{Model} & 
\textbf{DA} & 
\textbf{ACF} &
\textbf{CPR} & 
\textbf{CRA} & 
\textbf{AC} &
\textbf{ODS}\\ 
\hline

 {Sepczuk et al.} (2018) 
\cite{sepczuk2018new}  & Contextual & Mathematical Rule-based & \texttimes & \texttimes & \texttimes & \texttimes & \checkmark & \texttimes \\

 {Wang et al.} (2019) 
\cite{wang2019context}  & Behavioral Biometrics & OC-SVM & \texttimes & \checkmark & \texttimes & \texttimes &\texttimes & \checkmark \\

 {Solano et al.} (2019) 
\cite{solano2019risk}  & Contextual \& Behavioral & Random Forest Classifiers & \checkmark & \checkmark & \texttimes & \texttimes & \checkmark & \texttimes \\

 {Ghosh et al.} (2019)
\cite{ghosh2019softauthz}  & Contextual & Statistical Model & \texttimes & \texttimes & \texttimes & \checkmark & \checkmark & \texttimes \\

 {Rivera et al.} (2020)
\cite{rivera2020risk}  & Contextual & Multi ML Models & \texttimes & \checkmark & \checkmark & \texttimes & \texttimes & \texttimes \\

 {Li et al.} (2020)
\cite{li2020wrist}  & Behavioral Biometrics & Random Forest Classifiers & \texttimes & \checkmark & \texttimes & \checkmark & \texttimes & \checkmark \\

 {Xu et al.} (2020) 
\cite{xu2020gait}  & Behavioral Biometrics & Multi ML Models & \texttimes & \checkmark & \texttimes & \checkmark & \texttimes & \checkmark \\

 {Wu et al.} (2020)
\cite{wu2020caiauth}  & Contextual \& Behavioral & Multi ML Models & \checkmark & \checkmark & \texttimes & \checkmark & \texttimes & \checkmark \\

 {Buriro et al.} (2021) 
\cite{buriro2021risk} & Contextual \& Behavioral & Statistical Model & \checkmark & \texttimes & \texttimes & \checkmark & \checkmark & \texttimes \\

 {Papaioannou et al.} (2022) 
\cite{papaioannou2022toward}  & Contextual \& Behavioral & Multi ML Models & \texttimes & \checkmark & \texttimes & \checkmark & \texttimes & \checkmark \\

 {Liu et al.} (2022) 
\cite{liu2022log} & Contextual \& Behavioral & Bayesian Inference & \checkmark & \checkmark & \checkmark & \texttimes & \texttimes & \texttimes \\

 {Singh et al.} (2022) 
\cite{singh2022resilient} & Contextual \& NIDS & Multi ML Models & \checkmark & \checkmark & \texttimes & \checkmark & \checkmark & \texttimes \\

 {Weifling et al.} (2022) 
\cite{wiefling2022pump} & Contextual & Logistic Regression & \checkmark & \checkmark & \checkmark & \texttimes & \texttimes & \texttimes \\

 {Picard et al.} (2023) 
\cite{picard2023rlauth} & Contextual & Reinforcement Learning & \checkmark & \checkmark & \texttimes & \checkmark & \checkmark & \checkmark \\

\textbf{F-RBA (Our Framework)} & Contextual \& Behavioral & FL-based Autoencoder & \checkmark & \checkmark & \checkmark & \checkmark & \checkmark & \checkmark \\
\hline
\end{tabular}}
\end{table*}

Advancements in RBA have significantly enhanced online security. Freeman et al. \cite{freeman2016you} introduced a risk engine using logistic regression to assess contextual information for additional login verification. Wiefling at al. \cite{wiefling2022pump} expanded on this with a long-term analysis of RBA on a large-scale online service. Their approach supports dynamic adjustment through regular retraining and combines generalization with personalization by using all user data in the training phase. In their system, high-cardinality categorical features are processed based on their frequency of occurrence and transformed using hash functions to create fixed-size numerical representations for efficient storage and retrieval via hash tables resulted in increased user data privacy. While these studies provide valuable insights for RBA optimization and adoption, several challenges remain. These include computational intensity for server-side training of large datasets, insufficient consideration of user asset criticality, and incapability of continuous authentication.

Recent research has explored diverse approaches to RBA. Server-centric solutions include a scheme by Buriro et al. \cite{buriro2021risk} employing behavioral biometrics for continuous verification; they used touch-timing and hand-movement gestures with multiple ML models to enhance accuracy. Singh et al. \cite{singh2022resilient} developed an adaptive system that adjusts based on ML-derived risk assessments, incorporating Network Intrusion Detection Systems and surpassing existing solutions in resilience. Liu et al. \cite{liu2022log} introduced a lightweight Bayesian model for authentication compromise detection that improves accuracy over time and addresses the cold start problem. However, these server-centric approaches face challenges related to single point of failure and user privacy risks. In contrast, an on-device approach, RLAuth \cite{picard2023rlauth}, uses deep reinforcement learning to provide continuous risk assessment and dynamically adjust the level of authentication challenge based on context. However, it remains vulnerable to attacks in similar contexts. While RLAuth represents some advancements in RBA, it still struggles to combine generalization with personalization for more accurate risk assessment and address cross-device profiling, highlighting areas for further improvement in RBA systems.

Our comprehensive review reveals that despite the evolution from rule-based to ML-driven RBA systems, current solutions struggle with some fundamental challenges: maintaining user privacy in distributed environments, handling heterogeneous data distributions across users, and enabling efficient cross-device risk assessment without compromising security. Although recent studies, such as \cite{liu2022log} and \cite{picard2023rlauth}, have made progress in continuous authentication and on-device processing, they still fall short of providing a unified solution that addresses all of these challenges simultaneously. Our proposed framework introduces a novel approach to these challenges through a combination of federated learning and innovative feature engineering. Table \ref{tab:Main-Comparison} provides a comprehensive overview of recent RBA systems, highlighting current limitations in the field, and positioning our framework's advancements.
To compare existing contributions, we identified key features of RBA systems through analysis of the related work. These key features include:
(1) Dynamic Adjustment: Dynamically updates risk evaluation based on monitoring user activities and environmental changes;
(2) Automatic Context Finding: Automatically identifies contextual and behavioral patterns using ML on historical user data;
(3) Collective Pattern Recognition: Aggregates individual user data or models into a global model, balancing general trends with personalized data for improved risk evaluation;
(4) Continuous Risk Assessment: Conducts real-time, ongoing risk evaluation throughout the user session, enabling prompt responses to suspicious activities or contextual changes;
(5) Asset Criticality: Considers the sensitivity or importance of accessed data or resources to adjust risk evaluation and authentication requirements; and
(6) On-Device Setup: Deploys the risk engine and stores user data locally on the device, enhancing privacy by transmitting only the calculated risk level to the server.

\subsection{Federated Learning in Cybersecurity}
\label{subsec:Federated Learning in Cybersecurity}

Federated Learning (FL) is a decentralized approach for training ML models across various devices while keeping data local \cite{zhang2021survey}. It enhances model accuracy and privacy by aggregating updates from local models to refine a global model \cite{wen2023survey}. Devices perform local computations and transmit only model updates, reducing computational load and improving security. FL has emerged as a powerful cybersecurity approach that addresses challenges in privacy \cite{zhao2020local}, intrusion detection \cite{jin2024fl}, and user authentication \cite{sousa2024fedaaa}. Despite its widespread adoption, FL has not been directly applied to risk evaluation engines in RBA systems. We examine recent FL-based anomaly detection techniques that could be applied to RBA, highlighting this unexplored area.

Anomalies in datasets are instances that exhibit statistically infrequent or unexpected attributes, quantifiable through features that enable similarity metric computations. Three primary categories are distinguished \cite{landauer2023deep}: (1) point anomalies in independent datasets; (2) anomalies in interdependent data; and (3) collective anomalies in dependent data. This classification facilitates detection methodologies across diverse data structures. Notably, FL-based approaches can potentially identify and integrate all three categories, providing a comprehensive view of anomalous patterns across various contexts and scales.

Recent advancements in FL have leveraged autoencoders for anomaly detection. Vucovich et al. \cite{vucovich2023anomaly} introduced FedSam for intrusion detection across heterogeneous clients. Li et al. \cite{li2023autoencoder} proposed an autoencoder-based method for streaming non-stationary data. Yadav et al. \cite{yadav2021unsupervised} implemented an unsupervised technique for IoT intrusion detection, achieving 97.75\% accuracy while preserving privacy. These approaches demonstrate the versatility of autoencoders in federated settings, effectively addressing data heterogeneity, privacy, and adaptability—all factors applicable to our risk assessment purposes.


\section{Overview and Objectives}
\label{sec:System Model and Facets}

This section provides the system model, outlines the framework's key assumptions, and discusses various evaluation facets of RBA systems addressed by our proposed framework. Our main focus is on the risk evaluation engine and user data collection, rather than specific authentication techniques or the handling of user credentials for authentication.

\subsection {System Model}
\label{subsec:System Model}

\begin{figure}[ht]
\centerline{\includegraphics[width=0.49\textwidth]{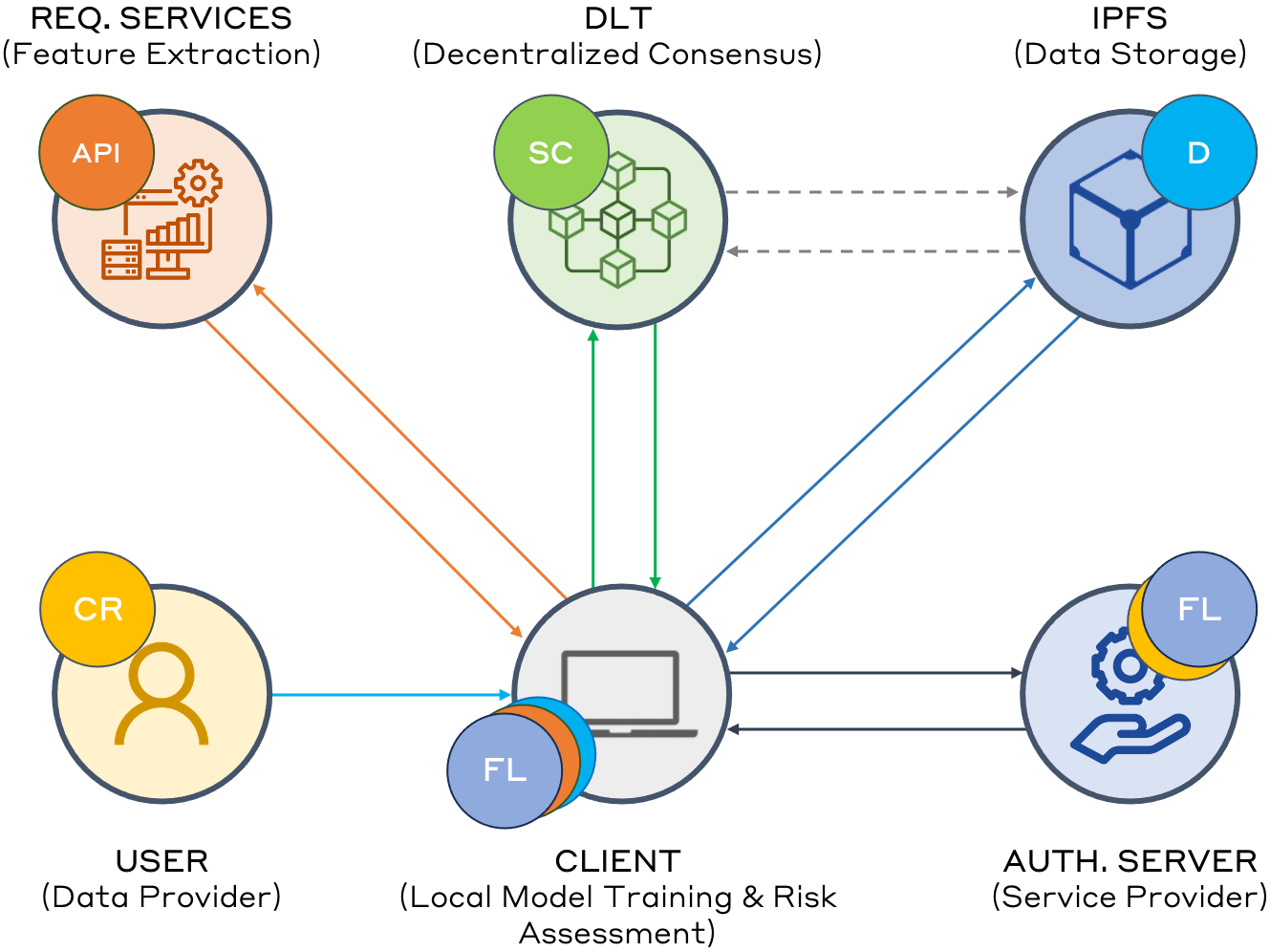}}
\caption{System Model of our Proposed Framework (F-RBA)}
\label{fig:System Model of our Proposed Framework (F-RBA)}
\vspace{-0.4cm}
\end{figure}

The proposed Federated Risk-Based Authentication (F-RBA) framework consists of six primary components: User, Client Application, Third-Party Services, an Authentication Server, IPFS, and Distributed Ledger Technology (DLT), as illustrated in Figure~\ref{fig:System Model of our Proposed Framework (F-RBA)}.

The \textit{User} is a registered entity who submits verifiable credentials (CR) through the client application during login. The \textit{Client Application} collects contextual data such as timestamps, IP addresses, and device types. It evaluates risk locally using a Federated Learning (FL) model, regularly retrains the local model, and communicates with the authentication server during both registration and login. \textit{Third-Party Services} provide secure Application Programming Interfaces (APIs), such as Cisco Talos \cite{cisco_talos_intelligence} and IPQS\cite{ip_quality_score}, offering essential data to the client application, including public IP details and IP reputation for more accurate risk-based authentication.
The \textit{Authentication Server} stores user credentials and plays a central role in the authentication process. It selects authentication methods based on assessed risk score, aggregates local model updates to maintain a global model, and grants user access accordingly. The \textit{IPFS} component is used for decentralized, off-chain storage of user login data, ensuring efficient retrieval, high availability, and generating Content Identifiers (CIDs) for stored data, which facilitates secure and traceable data management \cite{dwivedi2023smart}.
The \textit{DLT} records the CIDs of user data stored in IPFS and manages access control via Smart Contracts. It ensures secure recording and retrieval of aggregated user login data, preserving both privacy and transparency. Additionally, DLT supports secure data retrieval for risk assessment and model retraining within the F-RBA framework.

\subsection {Main Assumptions}
\label{subsec:Main Assumptions}

Our framework operates under several key assumptions concerning both the client application and the authentication server. We assume that all APIs involved are reliable in terms of availability, communication, and the provision of accurate and up-to-date data. The client application is expected to maintain security and reliability throughout its operations, including contextual data extraction, risk assessment model execution, and risk score generation. Additionally, all communications between the client application, the authentication server, and APIs are assumed to be secure and fully encrypted.
Furthermore, we assume that IPFS and DLT offer high speed and sufficient throughput in their respective roles. IPFS facilitates decentralized data storage and retrieval, while DLT efficiently records and manages data references, ensuring data integrity, security, and efficiency.
Regarding the authentication server, we assume that user credentials, such as passwords and security questions, are securely stored. The server is considered a reliable entity, ensuring high availability with no downtime for its authentication services, as well as maintaining security and reliability in global model aggregation.

\subsection {Evaluation Facets of RBA Systems}
\label{subsec:Evaluation Facets of RBA Systems}

In designing and developing an RBA system, several interrelated factors must be carefully balanced: performance, scalability, user experience, privacy, and security. Performance is critical for ensuring quick risk assessments and fast processing of login attempts and re-authentication with minimal latency. Scalability is equally important, enabling the system to efficiently handle a growing number of users and devices without degrading performance. User experience plays a pivotal role in reducing disruptions by triggering additional verification only for unusual activities, while adapting authentication methods based on the user's device and context. Privacy is ensured by protecting sensitive user data and limiting processing to essential information, thus maintaining confidentiality without compromising security. Lastly, security is fundamental to providing robust protection against unauthorized access and attacks, ensuring that the system remains secure and resilient.

Our proposed framework primarily enhances user data privacy and improves the scalability of training reliable models in large-scale environments. In addition, it offers significant security benefits for RBA, which we discuss in the following sections. These security enhancements are a valuable byproduct of our distributed privacy-centric approach, creating a more robust and resilient authentication system.


\section{Proposed Framework}
\label{sec:Proposed Framework}

In this section, we introduce the F-RBA framework, which utilizes a federated ML model for risk assessment. This framework is designed to enhance security in distributed environments. The F-RBA is a general-purpose, continous, context-aware RBA that considers both the current context and the user's historical data at the time of login.
First, we detail the risk assessment workflow of our proposed framework. Next, we describe the features used to evaluate the risk of each login attempt, considering the user's history. In addition, we explain our approach to feature engineering. These features capture the data essential for the framework's risk assessment. Finally, we present the FL model used to assess the risk of each login attempt, demonstrating how it leverages the contextual and historical data to make real-time risk evaluations.

\subsection{Overview of the Risk Assessment Workflow}
\label{subsec:Overview of the Risk Assessment Workflow}

\begin{figure}[ht]
\centerline{\includegraphics[width=0.49\textwidth]{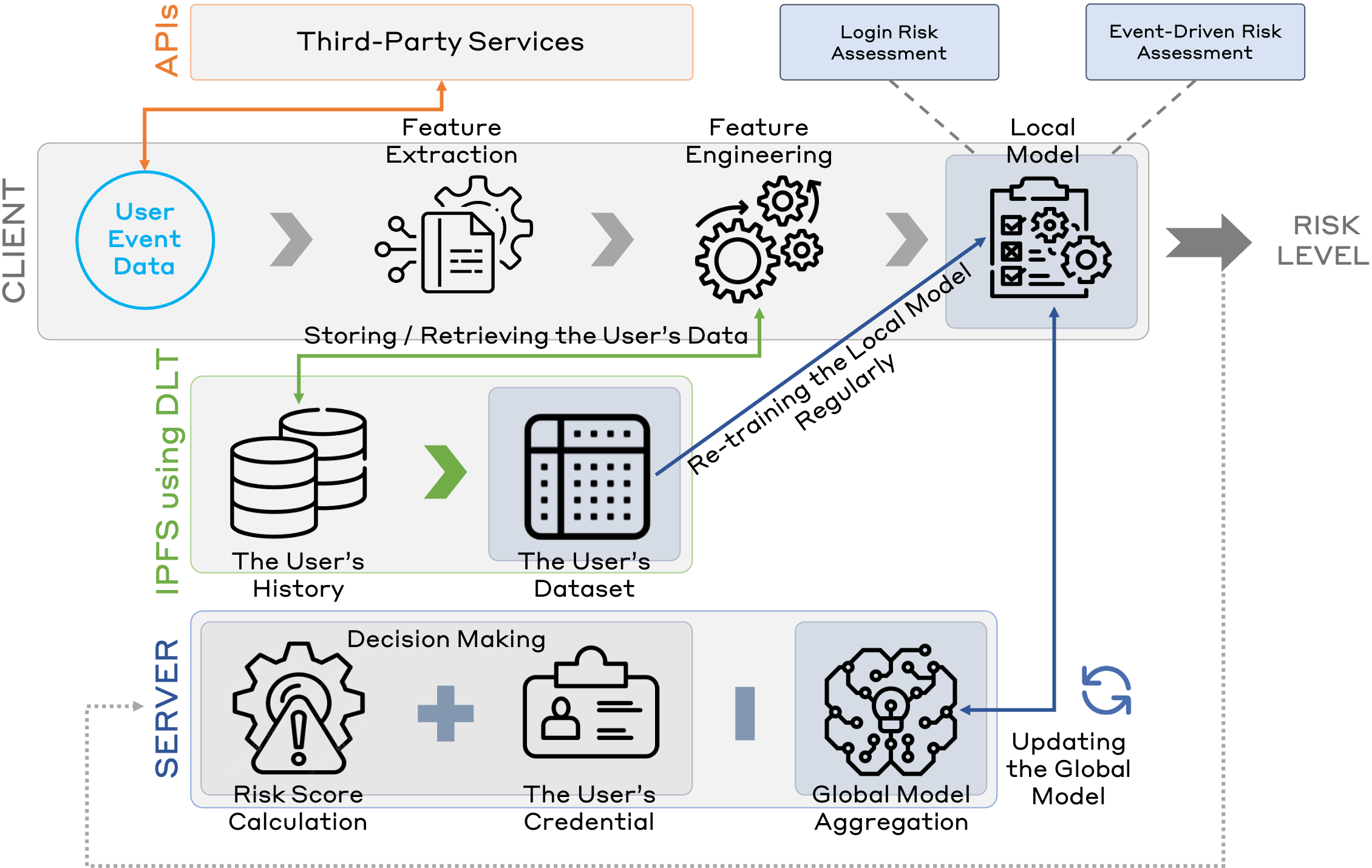}}
\caption{The Risk Assessment Workflow}
\label{fig:The Risk Assessment Workflow}
\vspace{-0.4cm}
\end{figure}

In this paper, we propose a framework that employs Horizontal Federated Learning (HFL) \cite{yang2020horizontal} where Clients with identical feature spaces but diverse sample spaces collaboratively train a global model on a central server, and a Semi-Synchronous Federated Learning (Semi-Sync FL) approach \cite{stripelis2022semi}, which balances efficiency and consistency by allowing flexible update timings with periodic synchronization. Our framework utilizes Federated Averaging (FedAvg) \cite{mcmahan2017communication} for effective model aggregation, integrating local model updates into a robust global model. This subsection illustrates the workflow of risk assessment and the risk evaluation model retraining (see Figure~\ref{fig:The Risk Assessment Workflow}) in our proposed framework. The components forming the Risk Assessment Workflow are the following:
(1) User Login Attempt: the user attempts to log in through the client application installed on their device(s).
(2) Data Collection: the client application collects the user's data, including system time zone, login timestamp, user agent string, and RTT (obtained by securely pinging the authentication server). Through third-party service APIs, it obtains the device's public IP address, geo-location data, and IP reputability.
(3) Feature Extraction: the client application extracts all features needed to assess risk from the user's data.
(4) Historical Data Check: the client application checks the user's historical data on the device. If outdated or missing (e.g., due to infrequent device use), it retrieves updated data from IPFS using the CIDs recorded on the Distributed Ledger Technology (DLT). The client uses a public-private key pair generated during registration to authenticate and encrypt communications with both IPFS and the DLT smart contract. Otherwise, it uses the existing complete history on the device.
(5) Login Risk Assessment: the client application performs feature engineering and uses the local model to assess risk level.
(6) Credentials and Risk Level Transmission: the client application sends the user's credentials (username and password) and the assessed risk level to the authentication server.
(7) Authentication Decision: based on the assessed risk score, calculated using a scoring grid (which we discuss in a following subsection) that incorporates the user's asset criticality, the authentication server may require additional authentication steps or, in extreme cases, block access.
(8) Data Storage: if the login is successful, the client encrypts the similarity-based format of the login record using the public key and stores the data in IPFS through a smart contract on the DLT, which manages the records of the user's history and dataset.
(9) Continuous Event-Driven Risk Assessment: During an active session, when users initiate events or transactions, the client application performs real-time risk assessment. The process evaluates the similarity between current and login-time contexts, while the authentication server makes decisions based on the risk score calculation (defined in subsection 4.6). This approach ensures continuous monitoring of user actions, enabling adaptive security responses when necessary. Through constant context validation, the system is capable of mitigating session hijacking attacks by detecting anomalous context patterns in real-time.

For the global model aggregation, the following actions are implemented:
(1) Dataset Threshold Check: when a user's dataset exceeds a certain number of new records, as tracked by a counter, the user device currently in use retrieves the user's dataset from IPFS using the CIDs obtained from the DLT.
(2) Local Model Retraining: the local model retrains, using the current global model weights as a starting point, to update itself with the user's complete dataset.
(3) Model Update Transmission: after retraining, the local model sends its updates and risk thresholds to the authentication server.
(4) Update Aggregation: the authentication server stores these local model updates and thresholds until a minimum number of contributors have sent their updates.
(5) Global Model Redistribution: the server averages the updates and thresholds, considering the weight of each contributor's dataset size, and redistributes the new model (with updated weights) to the clients.

In the following, we detail the process, from data extraction and model training to risk score calculation.

\subsection {Features}
\label{subsec:Features}

This paper introduces a risk engine that assesses the security risk of login attempts using contextual data. Our federated ML model analyzes features extracted from a base set (see Table~\ref{tab:Basic-Contextual-Features}), including:
(I) Login Timestamp: It identifies unusual access patterns.
(II) IP Address: It helps in geographical and network-based risk assessment.
(III) User Agent: It provides browser and operating system information, and device type.
(IV) System Identity: It provides time zone and language setup on the system.
(V) Connection Type: It shows the type of connection, such as WiFi or Cellular.
(VI) Round-Trip Time (RTT): It indicates network anomalies or geolocation discrepancies.
(VII) Successful Login: It is a binary indicator of login success.
(VIII) Is Benign IP: It is a binary feature that classifies IP addresses based on reputation and historical data.
These features form the foundation for our risk assessment model.
\begin{table}[ht]
\vspace{-0.2cm}
\centering
\caption{Basic Contextual Features}
\label{tab:Basic-Contextual-Features}
\fontsize{7}{8}\selectfont
\resizebox{\linewidth}{!}{ 
\begin{tabular}{p{0.9cm}p{3cm}p{1.4cm}p{3cm}}
\hline
\textbf{Code} & \textbf{Feature} & \textbf{Data Type} & \textbf{Range or Example} \\ \hline
I & Login Timestamp & String & 2020-02-03 16:19:21.693 \\
II & IP Address & String & 0.0.0.0 - 255.255.255.255 \\
III & User Agent (UA) & String & Mozilla/5.0... \\
IV & Time Zone \& Lang & String & "Asia/Tokyo", "en-US" \\
V & Connection Type & String & WiFi, Cellular, Ethernet \\
VI & Round-Trip Time (RTT) & Integer & 1-8600000(ms) \\
VII & Successful Login & Boolean & True (1) or False (0) \\
VIII & Is Benign IP & Boolean & True (1) or False (0) \\
\hline
\end{tabular}}
\vspace{-0.2cm}
\end{table}

\begin{table*}[ht]
\centering
\caption{Comprehensive Engineered Features}
\label{tab:Comprehensive-Engineered-Features}
\fontsize{7}{8}\selectfont
\resizebox{\textwidth}{!}{
\begin{tabular}{p{0.4cm}p{4cm}p{0.8cm}p{2.8cm}p{2.6cm}p{1.3cm}p{1.8cm}p{2.6cm}}
\hline
\textbf{\#} & \textbf{Feature} & \textbf{Code} & \textbf{Category} & \textbf{Subcategory} & \textbf{Data Type} & \textbf{Value Type} & \textbf{Range or Example} \\ \hline
1 & Hour of Day & I & Quantitative (Cyclic) & Temporal Data & Integer & Finite Set  &  Range of 0 to 23 \\
2 & Day of Week & I & Quantitative (Cyclic) & Temporal Data & Integer & Finite Set & Range of 0 to 6 \\
3 & Is Working Day & I & Qualitative (Categorical) & Temporal Data & Boolean & Finite Set & True (1) or False (0) \\
4 & Number of Logins Per Day & I & Quantitative (Binary) & Behavioral Metric & Integer & Finite List & True (1) or False (0) \\
5 & Time Between Logins & I & Quantitative (Linear) & Behavioral Metric & Integer & Continuous & 360000 (second) \\
6 & IP Address Range & II & Qualitative (Categorical) & Network Identifier & String & Unbounded Set & 8.8.8.*, 255.255.255.* \\
7 & Autonomous System Number (ASN) & II & Qualitative (Categorical) & Network Identifier & Integer & Unbounded Set & 100, 3000, 49051 \\
8 & Country, Region, City (3 Features) & II & Qualitative (Categorical) & Geo-Location Data & String & Unbounded Set & CA, Quebec, Montreal \\
9 & OS Name and Version & III & Qualitative (Categorical) & User Agent Information & String & Unbounded Set & Windows 11, iOS 17.6 \\
10 & Browser Name and Version & III & Qualitative (Categorical) & User Agent Information & String & Unbounded Set & Chrome 125.0.6422.142 \\
11 & Device Type & III & Qualitative (Categorical) & User Agent Information & String & Unbounded Set & Desktop, Tablet, Mobile \\
12 & Time Zone, Language (2 Features) & IV & Qualitative (Categorical) & System Identity & String & Unbounded Set & Asia/Tokyo, en-US \\
13 & Connection Type & V & Qualitative (Categorical) & Connectivity Metric & String & Finite Set & WiFi, Cellular, Ethernet \\
14 & Round-Trip Time (RTT) & VI & Quantitative (Linear) & Connectivity Metric & Integer & Continuous & 1 - 8600000 (ms) \\
15 & Unsuccessful Login Number & VII & Quantitative (Linear) & Behavioral Metric & Integer & An Integer & True (1) or False (0) \\
16 & Is Benign IP & VIII & Qualitative (Binary) & IP Reputation & Boolean & Only Two States & True (1) or False (0) \\
\hline
\end{tabular}}
\vspace{-0.4cm}
\end{table*}

Our risk assessment model leverages a diverse set of features, derived from basic login data. These include temporal factors (e.g., hour, day, login frequency, inter-login time), network metrics (e.g., IP range and RTT), geographical data, user agent details, authentication outcome, and IP reputation. Features are categorized by value types: unbounded sets (e.g., IP, OS name, and device type), finite sets (e.g., hour, day, device type), continuous variables (e.g., RTT and inter-login time), and binary features (e.g., IP reputation). Table~\ref{tab:Comprehensive-Engineered-Features} provides a comprehensive classification of extracted features, with codes indicating their base feature origins. This feature set enables sophisticated analysis of login patterns and potential threats. For example, temporal features identify unusual access times, Autonomous System Number (ASN) provides network insights, and user agent data offers device information. By combining these features, our model can detect anomalies like unexpected locations, unusual devices, suspicious network behaviors, or unusual user behavior, creating a robust framework for login risk assessment.

\subsection {Feature Engineering}
\label{subsec:Feature Engineering}

Feature engineering is critical for developing effective user behavior profiles in real-time anomaly detection \cite{khadivizand2020towards}, particularly in RBA systems. In FL, addressing the challenges of heterogeneity and Non-Independent and Identically Distributed (Non-IID) data is essential for model convergence and performance. We propose an approach that emphasizes similarity measures over direct feature values to mitigate these issues.

In real-world RBA systems, high-cardinality categorical data such as IP addresses, ASNs, and user agents are processed based on their occurrence frequency. While this frequency-based approach proves effective in centralized systems, it faces challenges in FL environments since each client has only local visibility of feature occurrences across the distributed system. In ML, common categorical encoding methods have limitations: one-hot encoding suffers from dimensionality explosion with high-cardinality features, while ordinal encoding introduces false numerical relationships between categories. The complexity intensifies in FL settings where clients exhibit significantly different feature distributions. Local feature occurrences vary across geographical locations and device types, leading to non-overlapping feature spaces between clients. This heterogeneity across clients causes model divergence, reduced performance, and inconsistent feature representations.

To address these challenges and ensure consistent feature representation, we have developed a systematic feature engineering strategy. This strategy reduces heterogeneity and enables effective federated model training. We categorize login features into two primary groups:
\begin{enumerate}[wide, font=\bfseries, labelwidth=!, labelindent=0pt]
\item \textbf{Qualitative Features:} These are descriptive and categorical features that require specific processing due to their non-numeric nature:
\begin{itemize} 
\item \textit{Binary features}: Two-state (0 or 1) characteristics based on anomaly detection, such as \textit{Is IP Benign}, which returns an anomaly score based on current IP status from trusted security services (e.g., Cisco Talos). For these features, a score of 1 indicates normal/benign status and 0 indicates anomalous/suspicious status.
\item \textit{Categorical features}: Non-numerical characteristics (e.g., \textit{ASN}, \textit{IP Address Range}, \textit{Country}, \textit{Time Zone}, and \textit{Device Type}) represented by similarity to user's past context, not by specific values. This approach uses a 0-1 scale with 1 indicating complete similarity to past patterns.
\end{itemize}
\item \textbf{Quantitative Features:}  These are numerical features that require statistical processing with specific mathematical relationships:
\begin{itemize} 
\item \textit{Binary features}: Features that generate binary scores through specific criteria. \textit{Number of Logins Per Day} belongs to this subcategory, generating similarity scores based on users' historical daily login patterns.
\item \textit{Linear features}: Numerical features with different scoring approaches, where \textit{Time Between Logins} and \textit{RTT} use similarity scores based on comparison to historical patterns, while \textit{Unsuccessful Login Number} employs an anomaly score based on consecutive login failures.
\item \textit{Cyclic features}: Circular time-based features like \textit{Hour of Day} and \textit{Day of Week}, using similarity scoring against historical patterns.
\end{itemize}
\end{enumerate}

Our systematic strategy primarily uses similarity-based scoring for most features, with a few exceptions that employ anomaly-based scoring. All features output values on a unified 0-to-1 scale, where 0 indicates either the highest similarity to past patterns (for similarity-based features) or normal/expected behavior (for anomaly-based features), and 1 indicates either the lowest similarity to past patterns (for similarity-based features) or anomalous/suspicious behavior (for anomaly-based features). This combination of similarity and anomaly scoring, while maintaining a unified scale, helps reduce heterogeneity among clients and aligns feature representations. The alignment mitigates differences in data distributions caused by geographical or other factors, making data more homogeneous across clients. Consequently, it enhances the FL model's ability to aggregate and learn from diverse client datasets, improving convergence and performance even when local feature spaces inherently differ.

In the following sections, we provide a detailed analysis of the specific methods used to assess anomaly and similarity for qualitative and quantitative features, examining our approach to feature comparison and risk assessment.

\subsubsection{Qualitative Features}

In the context of analyzing user login attempts for risk evaluation, we indeed encounter qualitative or categorical features. It's important to note that within this group, we have two subgroups: binary and categorical.

For binary features, such as \textit{Is Benign IP}, we use a simple function that returns 1 if the condition is met and 0 otherwise as follows:
\begin{equation}
S_{\text{binary}}(x) = \begin{cases}
1 & \text{if } x \text{ is benign} \\
0 & \text{otherwise}
\end{cases}
\end{equation}

In this equation, $S_{\text{binary}}(x)$ is the anomaly score for the binary feature, and $x$ is the actual IP Address being evaluated. This straightforward approach provides a clear representation of binary states without requiring any updating over time.

For categorical features like \textit{IP Address Range}, \textit{ASN}, \textit{Country}, \textit{Region}, \textit{City}, \textit{Is Working Day}, \textit{OS Name and Version}, \textit{Browser Name and Version}, \textit{Device Type}, \textit{Time Zone}, \textit{Language}, and \textit{Connection Type}, we use a weight-based proportional approach. The similarity score for an item value is calculated as the ratio of the weight of that item value to the sum of weights for all items in that category:
\begin{equation}
S_{\text{cat}}(v) = \frac{W_v}{\sum_{i} W_i}
\end{equation}
\noindent where $S_{\text{cat}}(v)$ is the similarity score for item value $v$, $W_v$ is the weight associated with item value $v$, and $\sum_{i} W_i$ is the sum of weights for all items' values.
Importantly, to implement this similarity measure efficiently, we only need to store two key pieces of information: the time of the last update ($t_{last}$) for the entire set and the set of items ($S$) with their respective weights. This minimal data storage approach allows for efficient computation and updates of the weight-based proportions; this makes it suitable for real-time analysis of user context patterns in risk evaluation.

\subsubsection{Quantitative Features}

In analyzing user context for evaluating the risk of a user's current login attempt, we encounter three main subgroups of features: binary features, linear quantitative features, and cyclic features. These features play a crucial role in identifying patterns and anomalies in user activities.

Binary features in our system include \textit{Number of Logins Per Day}. This feature produces discrete outcomes and is directly applied to detect anomalies and threshold breaches in user behavior.

For \textit{Number of Logins Per Day}, we use an Interquartile Range (IQR) based outlier detection method \cite{vinutha2018detection}. This approach identifies unusual activity by comparing the current day's login count to historical patterns. The IQR, calculated as the difference between the third (75th percentile) and first (25th percentile) quartiles of historical daily login counts, serves as our measure of statistical dispersion.
\begin{equation}
\text{IQR} = Q_3 - Q_1
\end{equation}

We calculate quartiles by sorting the last $i$ login days' counts (e.g., 100 days) in ascending order. $Q_1$ is the value at the $\frac{n+1}{4}$th position, and $Q_3$ at the $\frac{3(n+1)}{4}$th position, where $n$ is the number of historical daily login counts. Using the IQR, we define our similarity score for the login count as follows:
\begin{equation}
S_{\text{count}}(n) = \begin{cases}
1 & \text{if } n \leq Q_3 + 1.5 \cdot \text{IQR} \\
0 & \text{otherwise}
\end{cases}
\end{equation}

\noindent where $S_{\text{count}}(n)$ is the similarity score for the current day's login count $n$. It returns 1 if $n$ is within the expected range based on historical data, and 0 if it's an outlier. An outlier is defined as any value exceeding $Q_3 + 1.5 \cdot \text{IQR}$, a common statistical threshold for extreme values. This approach effectively detects unusual login activity while considering the user's typical behavior patterns.

Linear quantitative features in our system include RTT, time between logins, and the number of consecutive unsuccessful logins. These features are continuous variables that necessitate more complex similarity calculations. To address this, we propose a unified centroid-based approach for comparing such continuous variables.

For linear quantitative features like \textit{RTT}, we employ an exponential moving average (EMA) approach. The similarity score is computed using a Gaussian function \cite{huang2021gaussian}:
\begin{equation}
S_{\text{linear}}(x) = e^{-\frac{1}{2} \left(\frac{x - \mu}{\sigma}\right)^2}
\end{equation}

In this equation, $S_{\text{linear}}(x)$ is the similarity score for the linear feature, $x$ is the current value of the feature, $\mu$ is the EMA of the feature, and $\sigma$ is the standard deviation estimate of the feature. This Gaussian function measures how close the current value is to the historical average, normalized by the standard deviation. It provides a smooth, continuous similarity score that peaks at the mean and decreases as values deviate from the mean.

For the \textit{Time Between Logins}, we use a log-normal distribution approach that incorporates the standard deviation of log-transformed data. The similarity score is calculated as:
\begin{equation}
S_{\text{time}}(\Delta t) = e^{-0.5 * \left(\frac{\ln(\max(\Delta t, 1)) - \mu_{\text{log}}}{\sigma_{\text{log}}}\right)^2}
\end{equation}
\noindent where $S_{\text{time}}(\Delta t)$ is the similarity score for the time between the current and the previous logins, $\Delta t$ is the current time difference between logins, $\mu_{\text{log}}$ is the EMA of log-transformed historical time differences, and $\sigma_{\text{log}}$ is the standard deviation of log-transformed historical time differences. This equation measures how similar the current login interval is to historical intervals, using a log transformation to handle the typically right-skewed distribution of time intervals (e.g., minutes, hours, days, or weeks). 

The \textit{Unsuccessful Login Number} feature detects potential security threats by monitoring consecutive unsuccessful login attempts. This feature provides a graduated response to unsuccessful login attempts, with the score decreasing linearly until it reaches zero after a fixed number of consecutive failures. Let $N$ be the fixed threshold for the number of unsuccessful login attempts (e.g., $N = 5$), and let  $h = [h_1, h_2, ..., h_k]$ represent the sequence of recent login attempts for a user account, where $h_1$ is the most recent attempt prior to the current one. We define $f(h_i)$ as an indicator function for each login attempt:
\begin{equation}
f(h_i) = \begin{cases}
1 & \text{if login attempt } h_i \text{ was unsuccessful} \\
0 & \text{if login attempt } h_i \text{ was successful}
\end{cases}
\end{equation}

Let $m$ be the number of consecutive unsuccessful logins immediately preceding the current attempt:
\begin{equation}
m = \max \{k : \sum_{i=1}^{k} f(h_i) = k\}
\end{equation}

Then, the \textit{Unsuccessful Login Number} feature score $S_{\text{unsuccessful}}(h)$ is defined as follows:
\begin{equation}
S_{\text{unsuccessful}}(h) = \max\left(0, 1 - \frac{m}{N}\right)
\end{equation}

A score of 0 indicates an anomaly ($N$ or more consecutive unsuccessful attempts), while scores closer to 1 represent more normal behavior. This feature is evaluated at the beginning of each new login attempt and resets to 1 when a successful login occurs.

The \textit{Unsuccessful Login Number} feature aids in detecting potential brute-force attacks, forgotten credentials, or unauthorized access attempts. By focusing on recent login history and providing a graduated response, it offers a responsive measure of account access patterns, enabling quick detection and response to suspicious activity.

Cyclic features in our system include time-based variables such as \textit{Hour of Day} and \textit{Day of Week}. These features have a circular nature and require a special approach to calculate similarity, especially considering that users may have multiple typical periods of activity within a cycle. To capture this complexity without storing individual data points, we use a fixed-size circular histogram approach. For each cyclic feature, we store:
\begin{equation}
H = [w_1, w_2, ..., w_n]
\end{equation}

\noindent where $n$ is the number of bins, and each $w_i$ represents the weight (counts of occurrences) of the $i$-th bin. Specifically, for the day of week feature, we use 7 bins (one for each day), and for the hour of day feature, we use 24 bins (one for each hour). This approach allows us to maintain a fixed storage size regardless of the number of data points processed. To calculate the similarity score for a new value $x$, we use a weighted circular distance equation:
\begin{equation}
S_{\text{cyclic}}(x) = \frac{1}{2}\left(\frac{\sum_{i=1}^n w_i \cdot \cos(\theta_x - \theta_i)}{\sum_{i=1}^n w_i} + 1\right)
\end{equation}

In this equation, $\theta_x = 2\pi x / \text{period}$ is the angular representation of the current value, $\theta_i = 2\pi i / n$ is the angular representation of the center of bin $i$, and $\text{period}$ is the cycle length (7 for days of the week, 24 for hours of the day). The addition of 1 and division by 2 normalizes the cosine output from [-1,1] to [0,1], providing a similarity score where 1 indicates the highest similarity.

This method uses cosine similarity to create a smooth gradient, ensuring closer cyclic values have higher similarity scores. It recognizes time's circular nature, understanding hour 1 is nearer to 23 than 8. It accurately captures proximity to activity peaks; with a 9 AM peak, 8 AM and 10 AM score higher than 6 AM, as similarity gradually decreases away from the peak. This approach correctly identifies times close to activity peaks, regardless of numerical representation. By using this method, we effectively capture complex patterns in cyclic features, including multiple behavior peaks, while maintaining minimal storage requirements. The fixed-size histogram represents sophisticated activity patterns without storing individual historical data points.

Categorical and cyclic features in our system use an exponential declining window to update their statistics. This declining window, implemented through a decay factor, is applied only to historical data and not to the current day's data. The decay factor is applied once when transitioning to a new day. For a generic statistic $S$, the update equation when a new day starts is:
\begin{equation}
S_{\text{decayed}} = \alpha S_{\text{old}}
\end{equation}

\noindent where $\alpha$ is the decay factor (e.g., 0.95) and $S_{\text{old}}$ is the value of the statistic from the previous day. After applying this decay, new data for the current day is incorporated without further decay:
\begin{equation}
S_{\text{new}} = S_{\text{decayed}} + S_{\text{current:day}}
\end{equation}

This approach prioritizes recent behavior while considering historical patterns. For categorical features specifically, the weights for each category value $v$ are updated as $W_v^{\text{new}} = \alpha W_v^{\text{old}} + I_v$, where $I_v$ is the indicator function for the current day's observations of value $v$.
This unified decay method balances adaptability to changing behavior with preservation of past patterns. By applying decay only to historical data and incorporating current data separately, the system effectively captures both long-term trends and recent shifts in behavior. For categorical features, a minimum weight threshold $m$ is established; items with weights below $m$ (e.g., 0.5) are eliminated, optimizing storage by preventing the retention of negligible items.

To our knowledge, only successful logins update the user's history, providing deterministic evidence of identity and preventing system poisoning by malicious actors. Features marked with the subcategory of behavioral metric in Table~\ref{tab:Comprehensive-Engineered-Features} (e.g., \textit{Number of Logins Per Day}) use a temporary variable updated by unsuccessful logins. This temporary variable doesn't update the user's history; only successful logins contribute to the historical record.

Heterogeneous data poses a significant challenge in FL, often causing model divergence, particularly when using complex neural networks. Current solutions to heterogeneity and non-IID problems in FL fall into three major categories: data-based, algorithm-based, and system-based approaches \cite{zhu2021federated}. Taking a data-based approach, we tackle this challenge through feature engineering techniques that transform the data to approximate IID and homogeneous conditions by converting raw features into similarity or anomaly scores, mapping each client's features into a [0, 1] range. This unified approach, applicable to both qualitative and quantitative features regardless of their original distribution or scale, aligns the feature space across clients. By focusing on similarity to past context or behavior rather than absolute values, we achieve more statistical consistency, reducing skewness, kurtosis differences, and feature heterogeneity across clients. The resulting homogeneity in data distributions enhances the FL model's ability to aggregate and learn from diverse datasets, promoting robust distributed learning.

Moreover, this feature engineering approach is efficient in practice. Using summary statistics, moving averages, and fixed-size histograms, we capture key user behavior patterns without storing extensive raw data. This minimizes storage requirements and enables quick data transfer, allowing fast computation of similarity and anomaly scores during logins, and contributing to a responsive risk evaluation system.

\subsection {Local Model}
\label{subsec:Local Model}

\begin{figure}[ht]
\centerline{\includegraphics[width=0.49\textwidth]{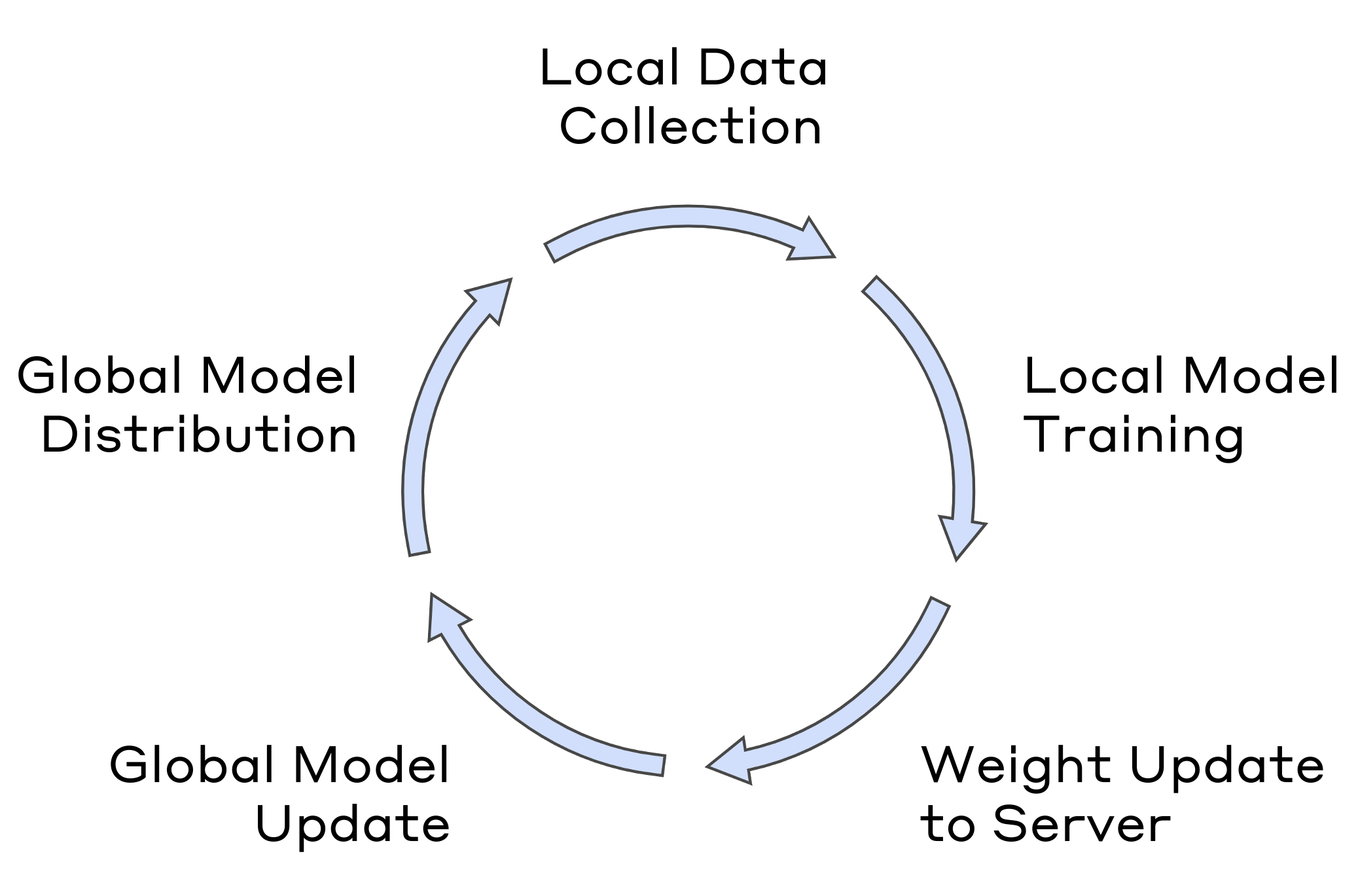}}
\caption{Local Model Update Cycle}
\label{fig:Global Cycle}
\end{figure}

FL offers an innovative solution for our authentication framework, processing users' contextual and behavioral data locally, similar to modern biometric systems. This method enhances user data privacy by providing only the risk level to the authentication server.
The continuous learning nature of FL ensures new user behaviors and contexts are considered, allowing fine-tuned risk evaluation for each access attempt. Our framework employs a dual-component system: local models on user devices and a global model aggregation on the authentication server, prioritizing privacy while leveraging collective learning.
We utilize a semi-synchronized strategy. Client applications trigger local model updates upon reaching a predefined threshold (e.g., 100 new login records). The local model retrains using user data and sends weights to the server. The global model updates occur when a significant portion of users (e.g., 20\%) contribute local updates, balancing timely updates with aggregated contributions. The updated global model is then distributed to all clients for their next training cycle (see Figure~\ref{fig:Global Cycle}).

The local model, an autoencoder (see Algorithm~\ref{algo:Local Model Training (Autoencoder)}), trains on a portion of the user's dataset in batches. For each data point, it performs a forward pass, computes loss with L2 regularization \cite{yang2023deep}, and updates parameters based on calculated gradients. The choice of autoencoder architecture is motivated by its ability to effectively learn compressed representations of normal behavior patterns while being robust to different data distributions across users. The model's encoder-decoder structure naturally handles the varying dimensionality of user contextual and behavioral patterns, making it particularly suitable for our authentication context \cite{ieracitano2020novel}.

\begin{algorithm}
\caption{Local Model Training (Autoencoder)}
\label{algo:Local Model Training (Autoencoder)}
\fontsize{8}{9}\selectfont
\begin{algorithmic}[1]
\REQUIRE User dataset $D$, batch size $b$, number of epochs $E$, learning rate $\eta$, L2 regularization parameter $\lambda$, update threshold $T$ (e.g., 64 new login records), and fraction of dataset to use $f$ (e.g., 0.1)
\ENSURE Trained local model weights $w$
\STATE Initialize model parameters $w$
\WHILE{new login records < $T$}
    \STATE Collect new login records
\ENDWHILE
\STATE $D_{\text{train}} \gets$ RandomSample($D$, fraction=$f$)
\FOR{epoch $= 1$ to $E$}
    \FOR{each batch $B \subset D_{\text{train}}$ of size $b$}
        \FOR{each data point $x \in B$}
            \STATE $\hat{x} \gets \text{Encoder}(x)$
            \STATE $x' \gets \text{Decoder}(\hat{x})$
            \STATE $L \gets \text{MSE}(x, x') + \lambda \|w\|_2^2$
            \STATE Compute gradients $\nabla L$
            \STATE $w \gets w - \eta \nabla L$
        \ENDFOR
    \ENDFOR
\ENDFOR
\STATE Transmit $w$ to authentication server
\STATE Return $w$
\end{algorithmic}
\end{algorithm}

The local model serves three critical functions in our system: (1) initial risk assessment during login attempts, evaluating the authenticity of user context at the point of authentication; (2) continuous event-driven risk assessment throughout the user session, monitoring for potential security anomalies in real-time; and (3) periodic local re-training to adapt to evolving user context and behavioral patterns while maintaining privacy guarantees.
A key feature is our standardized approach to hyperparameter management. We apply optimized hyperparameters across all local models, potentially accelerating learning and simplifying the process for end-users without requiring local model fine-tuning. This framework balances personalized risk assessment with collective learning, prioritizing user data privacy, security, and ease of use. Only valid registered users who have not opted out can contribute to global model updates, ensuring system integrity and fairness.

\subsection {Global Model Aggregation}
\label{subsec:Global Aggregation Model}

The global model aggregation follows the Federated Averaging (FedAvg) algorithm \cite{mcmahan2017communication, shrestha2024anomaly} (Algorithm~\ref{algo:Global Model Aggregation (FedProx)}), which consists of three main steps: server weight initialization, client local training, and server model aggregation and broadcast. In FedAvg, the server first initializes the global model weights and then distributes them to selected clients who perform local training on their private data. Finally, the server aggregates these locally trained models by averaging their parameters weighted by the size of each client's local dataset. To address the challenges posed by non-IID data distributions and to ensure proper model convergence, we employ FedProx \cite{li2020federated}, an extension of FedAvg that introduces client-side proximal regularization. The proximal term in the loss function is expressed as:

\begin{equation}
L_i(\mathbf{w}_u) = f_u(\mathbf{w}_u) + \underbrace{\frac{\mu}{2}\|\mathbf{w}_u - \mathbf{w}_g\|^2}_{\text{proximal term}}
\end{equation}

\noindent where $L_i(\mathbf{w}_u)$ represents the entire local loss function for user
$u$, including the proximal term. $f_u(\mathbf{w}_u)$ is the original loss function for user $u$, $\mathbf{w}_u$ represents the local model parameters, $\mathbf{w}_g$ is the global model parameters, and $\mu$ is the proximal term coefficient that controls the strength of regularization.

This proximal term penalizes large deviations between local and global models during client training, thus stabilizing the optimization process in heterogeneous data settings. When $\mu = 0$, FedProx reduces to the standard FedAvg algorithm, while the positive values of $\mu$ provide increasing levels of regularization, promoting consistency between local and global models. Furthermore, to ensure convergence and prevent unnecessary computations, we have considered a maximum iteration limit in our global model aggregation process.

\begin{algorithm}
\caption{Global Model Aggregation (FedProx)}
\label{algo:Global Model Aggregation (FedProx)}
\fontsize{8}{9}\selectfont
\begin{algorithmic}[1]
\REQUIRE Set of all users $U$, participation threshold $C$ (e.g., 10\% of user base), convergence threshold $\epsilon$, maximum iterations $max\_iter$, proximal regularization coefficient $\mu$, local learning rate $\eta$, number of local epochs $E$
\ENSURE Updated global model weights $w_g$
\STATE \textbf{Server initializes} global model weights $w_g$
\STATE Initialize empty user update set: $U_u \gets \emptyset$
\STATE $iter \gets 0$
\WHILE{$iter < max\_iter$}
\STATE Collect updates: $U_u \gets$ set of users who have uploaded updates
\IF{$|U_u| \geq C \cdot |U|$}
\STATE Store previous global weights: $w_{g\_old} \gets w_g$
\STATE \textbf{Server sends} current global weights $w_g$ and $\mu$ to all users in $U_u$
\FOR{each user $u \in U_u$ \textbf{in parallel}}
\STATE \textbf{User $u$ initializes} $w_u \gets w_g$
\STATE Get number of samples $n_u$ for user $u$
\FOR{each local epoch $e$ from $1$ to $E$}
\FOR{each batch $b$ in user $u$'s local data}
\STATE Compute gradient: $g \gets \nabla f_u(w_u; b) + \mu (w_u - w_g)$
\STATE Update local weights: $w_u \gets w_u - \eta \cdot g$
\ENDFOR
\ENDFOR
\STATE \textbf{User $u$ sends} updated weights $w_u$ and sample count $n_u$ to \textbf{Server}
\ENDFOR
\STATE Compute total samples: $N \gets \sum_{u \in U_u} n_u$
\STATE \textbf{Server aggregates} weights with sample-based weighting:
\STATE $w_g \gets \sum_{u \in U_u} \frac{n_u}{N} w_u$
\STATE Clear user update set: $U_u \gets \emptyset$
\STATE $iter \gets iter + 1$
\ENDIF
\ENDWHILE
\end{algorithmic}
\end{algorithm}
\vspace{-0.6cm}

\subsection{Risk Score Calculation}
\label{subsec:Risk Score Calculation}

The risk assessment process operates in two sequential phases: login-time and continuous event-driven assessment. It integrates two main factors: the configurable asset criticality ($AC$) and the risk level ($RL$), which is determined by a context-aware autoencoder model. The autoencoder quantifies deviation severity, with $RL \in \{0, 1, 2\}$, representing low, medium, and high risks, respectively.

Asset criticality is a configurable value set by the administrative team, represented as $AC \in \{1,2,3\}$, where $1$ represents the lowest, $2$ the medium, and $3$ the highest level of asset criticality. This limited number of categories based on asset impact is supported by existing regulations \cite{uk_gsc} and standards \cite{fips199}. Risk levels, on the other hand, are calculated during two phases: (1) during the login-time assessment, the model evaluates initial login context features, assigning the risk level ($RL_{\text{login}}$), and (2) after a successful login, in the continuous event-driven assessment phase, the model continuously monitors new events, comparing them with the login session's features and assessing the risk level ($RL_{\text{event}}$) accordingly.

The risk score calculation uses both the asset criticality and the determined risk levels to generate a base risk score ($BS$) ranging from 1 to 4. This calculation occurs on the authentication server, which handles user credentials and makes access control decisions. The base risk score is assigned as follows:

\begin{table}[!ht]
\centering
\fontsize{7}{8}\selectfont
\caption{The Risk Score Calculation Grid}\label{table:gird}
\vspace{-0.1cm}
\resizebox{\linewidth}{!}{ 
\begin{tabular}{>{\centering\arraybackslash}p{1.7cm} >{\centering\arraybackslash}p{1.7cm} >{\centering\arraybackslash}p{1.7cm} >{\centering\arraybackslash}p{1.7cm}}
\hline
\multirow{2}{*}{\textbf{Asset Criticality}} & \multicolumn{3}{c}{\textbf{Risk Level}} \\
\cmidrule(lr){2-4}
 & \textbf{0} & \textbf{1} & \textbf{2} \\
\hline
\textbf{1} & \cellcolor{green!25}1 & \cellcolor{green!25}1 & \cellcolor{yellow!35}2 \\
\textbf{2} & \cellcolor{green!25}1 & \cellcolor{yellow!35}2 & \cellcolor{orange!40}3 \\
\textbf{3} & \cellcolor{yellow!35}2 & \cellcolor{orange!40}3 & \cellcolor{red!45}4 \\
\hline\noalign{\smallskip}
\multicolumn{4}{c}{\textcolor{black}{
\fcolorbox{green!25}{green!25}{\rule{0pt}{2pt}\rule{2pt}{0pt}} No Risk \hspace{0.4cm} \fcolorbox{yellow!35}{yellow!35}{\rule{0pt}{2pt}\rule{2pt}{0pt}} Low Risk \hspace{0.4cm} \fcolorbox{orange!40}{orange!40}{\rule{0pt}{2pt}\rule{2pt}{0pt}} Moderate Risk \hspace{0.4cm} \fcolorbox{red!45}{red!45}{\rule{0pt}{2pt}\rule{2pt}{0pt}} High Risk}
} \\
\noalign{\smallskip}\hline
\end{tabular}}
\vspace{-0.1cm}
\end{table}

Table~\ref{table:gird} presents the risk score calculation grid for different AC levels. The base risk score is capped at 4 to maintain a clear distinction between standard risk levels (1-4) and the critical risk state (5) triggered by the security measure $S(t)$, which is calculated as follows:
\begin{equation}
S(t) = \frac{F(t)}{F_{\text{max}}} + \frac{H(t)}{H_{\text{max}}}
\end{equation}

\noindent where $F(t)$ is the number of consecutive failed login attempts right before the current login or within the current session, with $F_{\text{max}}$ (e.g., 5) being a standard limit for failed login attempts before a temporary lockout. $H(t)$ refers to the number of consecutive high-risk events (where $RL_{\text{event}}$ = 2), and $H_{\text{max}}$ (e.g., 3) reflects the threshold for significant high-risk events that indicate potential attack patterns. The dynamic threshold ($T_{AC}$) for each $AC$ level is defined as:
\begin{equation}
T_{AC} = 1 + \frac{n - AC}{n}
\end{equation}

\noindent where $n$ represents the maximum level of AC, equal to 3. This threshold varies based on the asset criticality level, being more lenient for lower criticality assets and stricter for higher criticality assets to ensure appropriate security measures. Finally, the risk score is determined by:
\begin{equation}
\text{Risk Score} = 
\begin{cases}
5 & \text{if } S(t) > T_{AC} \\
\text{$BS$} & \text{otherwise}
\end{cases}
\end{equation}

These scores can be interpreted as follows:
(1) No Risk: no security concern, normal situation (e.g., password is sufficient);
(2) Low Risk: minor monitoring is recommended (e.g., password and security questions are required);
(3) Moderate Risk: increased monitoring with preventive measures (e.g., password, security questions, and OTP are required);
(4) High Risk: immediate preventive actions are recommended (e.g., password, OTP, and email verification are required); and
(5) Critical Risk: severe situation, urgent intervention required (e.g., account is temporarily locked).


\section{Implementation and Analysis}
\label{sec:Implementation and Analysis}

In this section, we detail the implementation environment of our federated framework. We discuss the dataset employed, describe the local and global models, and explain the process of defining risk thresholds for three distinct levels. Subsequently, we evaluate the results and analyze the framework based on some of the key RBA facets outlined in the Overview and Objectives section.

\subsection{Environment}
\label{subsec:Environment}

The FL-based model prototype was implemented entirely in Python on a PC with an Intel(R) Core(TM) i9-9880H CPU @ 2.30GHz (8-core processor), 16 GB of RAM, and a 1TB SSD. A Conda environment on macOS was used to manage dependencies and isolate the project.
The autoencoder model was built using Keras running on TensorFlow. We utilized the FedAvg algorithm through TensorFlow Federated.
For feature extraction and engineering, the system employed Pandas, SciPy, and NumPy libraries. Scikit-learn was utilized for additional models, such as OC-SVM. Data visualization and graphical analyzes were performed using Matplotlib and Seaborn libraries.

\subsection{Dataset and Preprocessing}
\label{subsec:Dataset and Preprocessing}

One of the most significant limitations in this field is the scarcity of suitable datasets for context-based risk-based authentication. For our implementation, we identified the publicly available dataset \cite{wiefling2022pump} as the most appropriate, as it supports most of our features. However, it does not provide support for features IV and V, as listed in Table~\ref{tab:Basic-Contextual-Features}. We extracted relevant user data containing contextual information on login attempts from multiple users. From the 752 most recorded User IDs, we selected 500 users, excluding those with notable numbers of empty values for critical features (such as city and region) or predominantly unsuccessful logins. The selected users had record counts ranging from 200 to 6,417, totaling 185,420 records. This selection focused on users with more than 200 records to ensure meaningful sample sizes for the training phase. The first two users, with over 14 million and 70,028 records, respectively, were excluded to avoid imbalance. We cleaned and preprocessed the dataset to include only users with all necessary features, and applied the feature engineering techniques that subsequently normalized values to a 0-1 range for uniform representation across all records. Drawing on the statistics specific to each user's context in the dataset, along with the proposed feature engineering techniques, unexpected or significant deviations from typical contextual patterns are identified. Taking into account factors such as timing, IP address, location, and device characteristics, our approach detects anomalies in a user’s established contextual profile.

\subsection{Model Selection and Validation}
\label{subsec:Model Employed}

We employ the FedProx algorithm, an extension of FedAvg that uses sample-size-based weights to accommodate clients with varying dataset sizes and incorporates a proximal term to stabilize training. We utilize this aggregation algorithm over Federated Stochastic Gradient Descent (FedSGD) due to its intrinsic faster convergence and reduced network overhead \cite{mcmahan2017communication, li2020federated}. For each client, training is triggered upon accumulating 50 new successful login records. When triggered, the client selects its most recent $N = \min(500, \textit{total records})$ logins. For this dataset size, global aggregation occurs when 10\% of active users contribute their updates, ensuring consistent evolution of the model while maintaining system efficiency. We conducted 80 training rounds, monitoring the reconstruction loss of the global model. This number of training rounds is based on the setup assumptions, such as the number of new successful logins and the total number of records available.

\begin{figure}[ht]
\centerline{\includegraphics[width=0.49\textwidth]{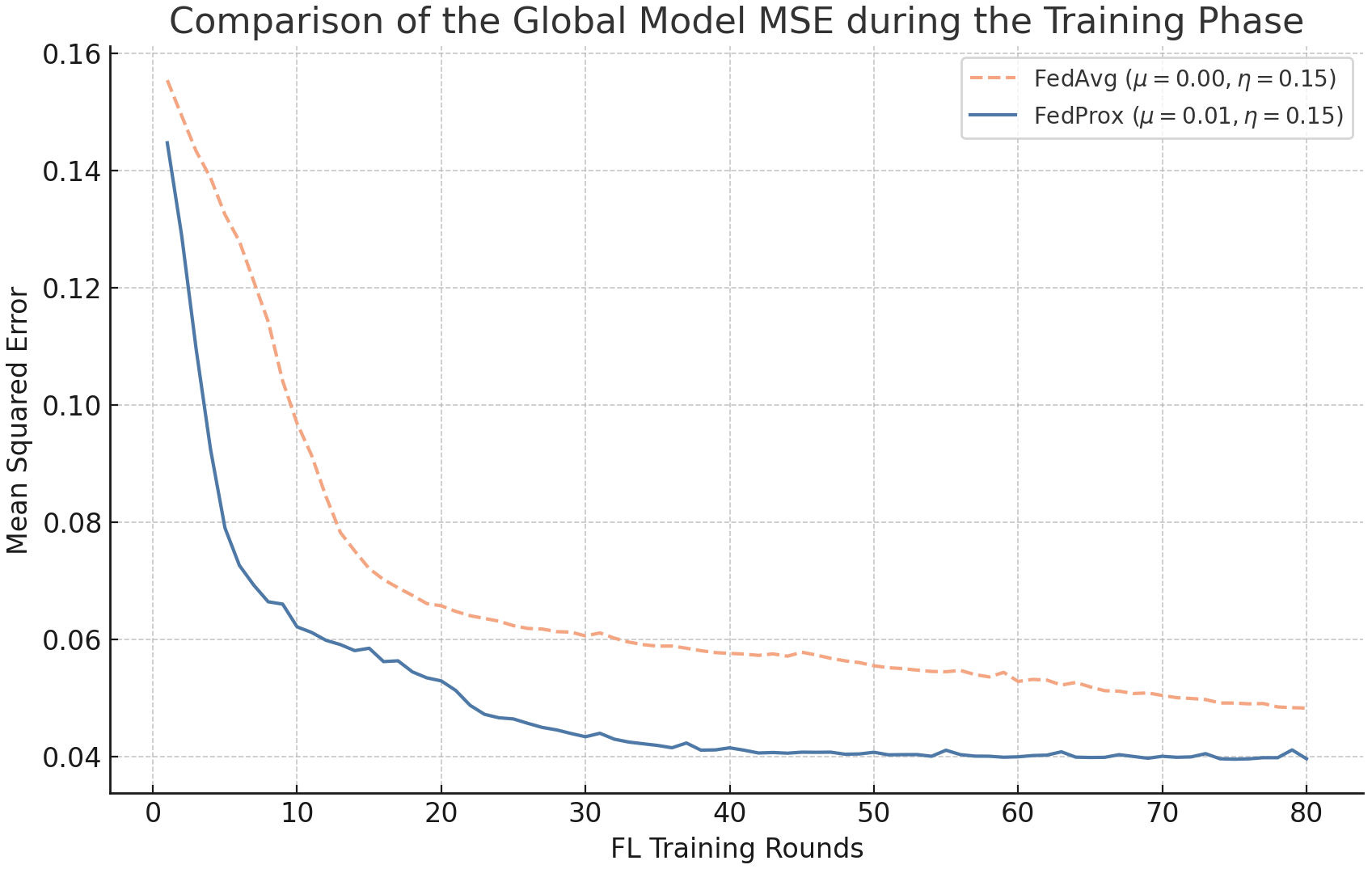}}
\caption{Mean Squared Error (MSE) with Different $\mu$}
\label{fig:Comparison of MSE}
\vspace{-0.4cm}
\end{figure}

Our symmetric deep autoencoder employs a 12-9-6 encoder and bottleneck architecture, complemented by a 9-12 decoder. The model utilizes Rectified Linear Unit (ReLU) activation functions for hidden layers. We apply L2 regularization ($\lambda = \text{\num{1e-4}}$) for FedAvg (where $\mu = 0.00$) and implement lightweight dropout layers to mitigate overfitting \cite{maheshwari2022autoencoder}, while employing a proximal term ($\mu = 0.01$) with a learning rate ($\eta = 0.15$) for FedProx, as illustrated in Figure~\ref{fig:Comparison of MSE}. FedAvg operates by having clients train locally and aggregate their updates through weighted averaging at the server, but struggles with client drift, non-IID data, and slow convergence. FedProx improves on this by adding a proximal term regularization that keeps client models from straying too far from the global model during training, making it more stable when dealing with heterogeneous data distributions while still using FedAvg's basic averaging mechanism. FedProx is particularly attractive because it implements these improvements entirely on the client side without adding any computational overhead to the server. The smooth decrease in MSE for FedAvg indicates that the non-IIDness is not severe, as we observe no spikes or sharp oscillations. However, FedProx demonstrates notable advantages with faster convergence and improved stability. This design and its parameters, optimized through heuristic experiments and grid search, effectively capture complex patterns while maintaining a focus on essential information. Keras-based random weight initialization ensures an unbiased start, addressing cold-start issues in the absence of a pre-trained base model. Although pre-trained models could accelerate convergence \cite{tan2022federated}, random initialization prevents potential bias \cite{guo2023fedbr} and promotes fairness in FL. For evaluation, we designate 20\% of users as the unseen test set and the remaining 80\% for training, allowing robust model assessment. Our model demonstrates consistent performance across diverse user patterns, with average reconstruction error varying by less than 9\% across different user subsets, indicating its effectiveness for heterogeneous user behaviors.

\subsection{Risk Level Thresholds}
\label{subsec:Risk Thresholds}

The process of identifying risk levels from reconstruction errors involves two main steps: finding optimal thresholds and classifying risks based on these thresholds. In our unsupervised model, we use Mean Squared Error (MSE) as the reconstruction error to determine risk levels. We investigated several thresholding methods. These include fixed percentiles, combinations of mean and standard deviation, and inter-quartile range.
Unsupervised models for anomaly detection \cite{landauer2023deep} often benefit from combining fine-tuned models with effective thresholding methods. We found that a combination of robust mean and Median Absolute Deviation (MAD) yields satisfying results, especially for users with a limited number of records. We determine two key thresholds using this robust statistical approach in conjunction with extreme value theory. First, we calculate an initial threshold and tail data as:
\begin{equation}
    \text{Initial threshold} = \mu_{\text{robust}} + 1.5  \cdot \text{MAD}
\end{equation}
\begin{equation}
    \begin{aligned}
        \text{Tail data} = \{ x - \text{initial threshold} \mid x \in \text{errors}, \\
        x > \text{initial threshold} \}
    \end{aligned}
\end{equation}

\noindent where, $\mu_{robust}$ is the robust mean calculated using the Minimum Covariance Determinant (MCD) estimator \cite{hubert2010minimum}, and MAD is the Median Absolute Deviation \cite{pham2001mean}. MCD and MAD estimators provide resilience against outliers that could skew threshold calculations. We then fit a Generalized Pareto Distribution (GPD) \cite{arnold2008pareto} to the tail data exceeding this initial threshold, ensuring the existence of a tail by including at least two records of the user's data. GPD was chosen for its effectiveness in modeling extreme deviations in user behavior patterns.

Subsequently, we apply the K-means clustering \cite{ahmed2020k} with $k=2$ (to distinguish between medium- and high-risk classes) to the full tail data to determine two thresholds: the lower threshold ($T_{\text{lower}}$), which equals the initial threshold, and the upper threshold ($T_{\text{upper}}$), which is determined by the K-means model on the full tail data. We define the risk level classification function as:
\begin{equation}
    \text{Risk Level} (x) = 
    \begin{cases}
        0 & \text{if } x \leq T_{\text{lower}} \\
        1 & \text{if } T_{\text{lower}} < x \leq T_{\text{upper}} \\
        2 & \text{if } x > T_{\text{upper}}
    \end{cases}
\end{equation}
\noindent where Risk Level($x$) determines the level of risk for the record $x$. Finally, we classify the risk levels based on these thresholds. Errors below or equal to the lower threshold are classified as normal (class 0), those between the lower and upper thresholds are classified as risky (class 1), and those exceeding the upper threshold are classified as high risk (class 2). For real-time anomaly detection, each user's local model should adaptively store the thresholds from the most recent training update to identify whether new records are anomalous. This approach enables adaptive personalized anomaly detection tailored to each user's unique login context patterns.

\subsection{Evaluation}
\label{subsec:Evaluation}

\begin{figure}[ht]
\centerline{\includegraphics[width=0.49\textwidth]{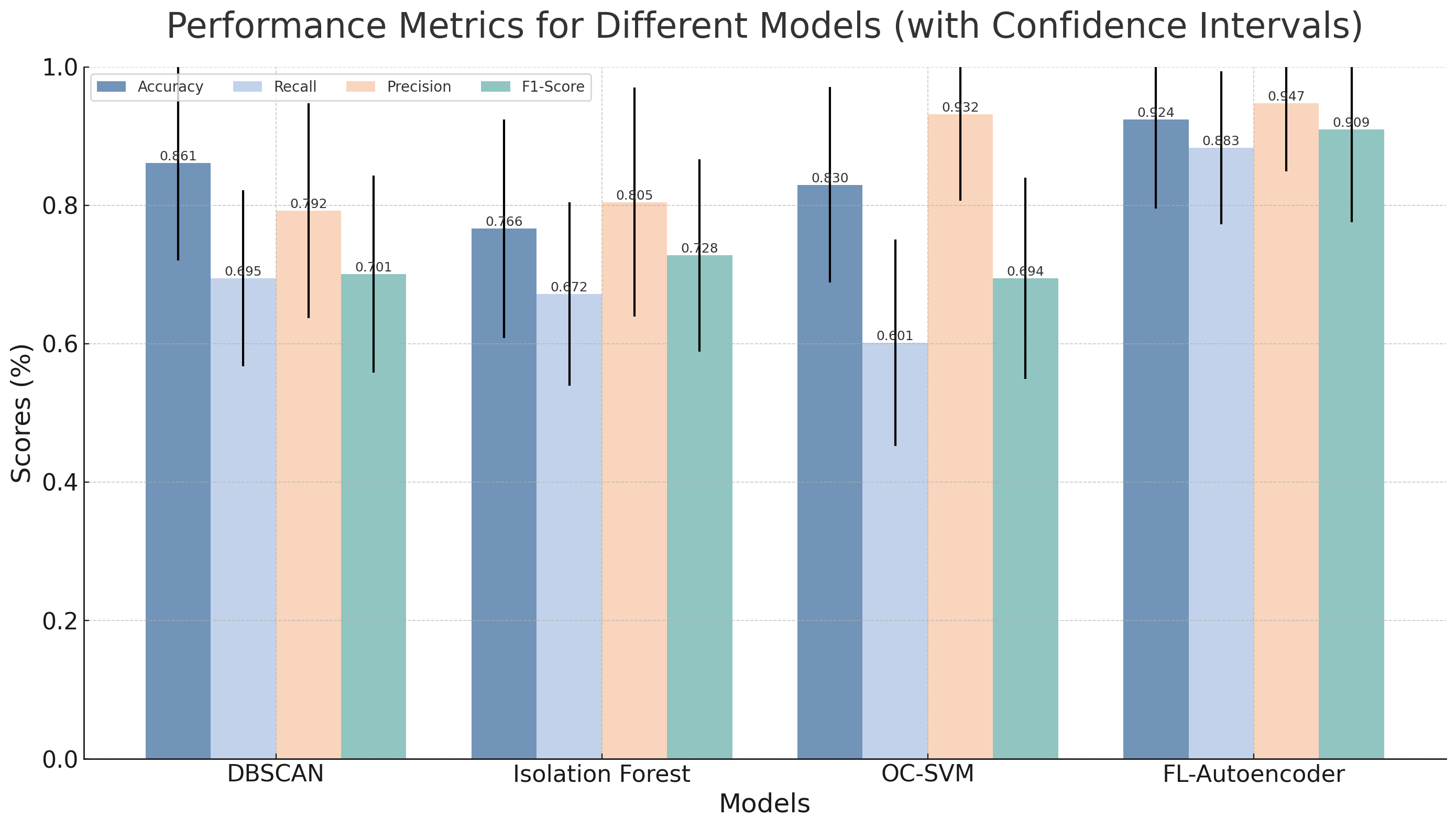}}
\caption{Comparative Performance Metrics Across Model Variants}
\label{fig:Comparative Performance Metrics Across Model Variants}
\vspace{-0.4cm}
\end{figure}

We evaluated our federated model against three conventional unsupervised models commonly used for anomaly detection \cite{samariya2023comprehensive} and in some related works: OC-SVM, Isolation Forest (IF), and Density-Based Spatial Clustering of Applications with Noise (DBSCAN). OC-SVM is commonly used for anomaly detection in high-dimensional data, DBSCAN is effective for detecting clusters of arbitrary shape and identifying outliers, and Isolation Forest is efficient for large datasets and designed specifically for anomaly detection. These models, not inherently suitable for FL, were trained and evaluated using the same datasets as our proposed model. All models were evaluated for their ability to perform binary classification, distinguishing between normal and anomalous login records.
We compared performance (see Figure~\ref{fig:Comparative Performance Metrics Across Model Variants}) based on four metrics \cite{hicks2022evaluation}: Accuracy, Recall, Precision, and F1-Score. Our model achieved an average True Positive Rate (Recall) around 88\%, demonstrating notable anomaly detection capability in login attempts. The federated model showed overall superior performance compared to other models, though not absolute superiority for all users' datasets. In particular, OC-SVM demonstrated relatively good accuracy and high precision but low recall, reflecting its high sensitivity to outliers. These results indicate that our proposed framework for distributing the risk evaluation engine is both feasible and performs well.

\renewcommand{\arraystretch}{1.2}
\begin{table*}[b]
\vspace{-0.2cm}
\caption{Comparison of F-RBA with the Most Recent Application-Agnostic, Context-Aware, ML-Based RBA Systems}
\vspace{-0.1cm}
\label{tab:Comparison-Privacy}
\centering
\small
\resizebox{\linewidth}{!}{
\begin{tabular}{p{4.5cm}p{9.8cm}C{1.4cm}C{1.4cm}C{1.4cm}C{1.4cm}C{1.4cm}}
\hline
\textbf{Contribution} & 
\textbf{Description} & 
\cite{liu2022log} &
\cite{singh2022resilient} &
\cite{wiefling2022pump} &
\cite{picard2023rlauth} &
F-RBA \\ 
\hline
Distributed Risk Evaluation & Cross-device client-based risk assessment, efficiently handling user growth. & \texttimes & \texttimes & \texttimes & \texttimes & \checkmark \\
Distributed Risk Model Training & Client-side model training for server overhead reduction. & \texttimes & \texttimes & \texttimes & \checkmark & \checkmark \\
Cold-Start Friendly & Accurate risk evaluation with limited user history for new users. & \checkmark & \texttimes & \checkmark & \texttimes & \checkmark \\
User Data Control \& Privacy & Login data protection against unauthorized access, leakage, and abuse. & \texttimes & \texttimes & \texttimes & \checkmark & \checkmark \\
\hline
\end{tabular}}
\vspace{-0.2cm}
\end{table*}

\subsection{Comparative Analysis}
\label{subsec:Comparitive Analysis}

This concise analysis highlights our framework's contributions to recent advancements in RBA systems, as demonstrated through our comparison in Table~\ref{tab:Comparison-Privacy} of the most recent application-agnostic, context-aware ML-based RBA frameworks. The table shows that our framework uniquely addresses scalability and privacy facets by providing a user-based on-device risk assessment and fully preserving user data privacy, which are not fully supported by other systems.

According to the most recent proposed RBA systems, even those such as \cite{picard2023rlauth}, which support on-device risk assessment engines, do not support a user-based on-device approach that considers the entire user profile across all devices. They only consider the history of the user on the current device. By utilizing this approach, we consider the whole contextual profile of the user at each login attempt, enhancing security and user experience.

Our framework addresses key challenges in RBA by balancing generalization and personalization. While consistent patterns in time, location, and device usage are typically normal, exceptions exist for specific user roles or lifestyles. For example, a traveling salesperson may frequently change locations, or a web developer might log in from various devices and browsers at different times. By considering both global norms and individual habits, our framework enhances anomaly detection accuracy. Our feature engineering reduces data heterogeneity among clients, addressing the challenge of heterogeneous data in FL and improving model convergence and performance.

We prioritize data privacy by protecting user login data from unauthorized access and empowering users to manage their personal data, including access and usage rights. RBA systems, such as \cite{picard2023rlauth}, provide this level of data privacy to users, preventing login data (whether raw or hashed) from being passed to or stored in authentication servers or by other parties outside the user's access and control. Although \cite{wiefling2022pump} employs hashing techniques to protect stored user data, it cannot eliminate all statistical inference risks. Moreover, users lack exclusive control over their data, leaving potential vulnerabilities to leakage and other security threats. Our framework further strengthens privacy through the use of FL and decentralized data storage, ensuring that sensitive user data remains on the user's device.

By utilizing FL, our framework overcomes the cold-start challenge, enabling reliable risk evaluation for new users or devices with limited historical data. Other RBA systems, such as those proposed by \cite{liu2022log} and \cite{wiefling2022pump}, which train their main models on all users' data, also provide this capability. However, all RBA systems utilizing collective learning—whether in a centralized setting (like \cite{liu2022log} and \cite{wiefling2022pump}) or a federated setting (like F-RBA)—are prone to model poisoning attacks. Groups of malicious actors can poison the model. To address this vulnerability, we will extend our work as discussed in the following section. Our framework distributes both risk evaluation and model training across client devices, potentially offering scalability for growing user bases and authentication requests while maintaining robust performance. Additionally, it supports broad applicability through similarity-based data formats while enabling custom access control, including time- or location-based restrictions for specific use cases.

\subsection{Data Transfer and Computation Analysis}
\label{subsec:Computation Analysis}

In our framework, data access and transfer overhead refers to the time required for data retrieval from decentralized storage and network transmission, while computational overhead represents the time for devices to process features and perform risk assessments for authentication. Our analysis assumes a low-end Wi-Fi Internet speed of 16 Mbps, with an error margin of 20\% for packet loss, network congestion, and speed variations, as supported by real-world network performance research \cite{macmillan2023comparative}.
Data packets of 500 kilobytes (kB) can be retrieved in approximately 300 milliseconds (ms). This 500 kB threshold was established through experimental analysis of user historical data, which showed maximum sizes of 250 kB per user, with the doubled capacity providing adequate headroom for various scenarios. The retrieval overhead stems primarily from IPFS operations (node discovery, content routing, and data transfer), while DLT overhead (Smart Contract query and CID lookup) remains constant regardless of data size. In optimized real-world IPFS setups, retrieving 500 kB typically takes around 250 ms, with worst-case scenarios remaining under 2 seconds \cite{trautwein2022design}. The client retrieves lightweight IP data from enterprise-scale APIs with approximately 100 ms latency, executing these requests in parallel with historical data fetching to minimize overall response time.

For computational overhead, the analysis encompasses both feature engineering and risk assessment times. The measurements incorporate the current feature space and model. Over 20 experimental runs (ensuring statistical robustness) the testing of the local model for new login—including feature extraction, feature engineering, and risk assessment utilizing the local model and comparing the reconstruction error with the thresholds—takes an average of 5.82 ms.

Our framework utilizes a DLT-based smart contract to manage IPFS-stored encrypted login data, with user-specific private keys enabling secure decryption and retrieval. This architecture enhances security by integrating usernames with the smart contract for controlled access, while maintaining cloud storage as a viable alternative. DLTs like Hedera \cite{baird2019hedera}, which can be deployed privately to reduce costs and customize architecture, achieve rapid transaction finality. The retrieval process is notably faster than storage due to reduced consensus overhead and IPFS's content-addressing capabilities. Risk assessment is achieved with sub-second response times, considering computational, data transfer, and encryption overhead factors. These performance metrics are consistently maintained under standard operating conditions, which include a high-speed Internet connection, an optimized IPFS architecture, and a fast DLT implementation running on a reliable and efficient infrastructure backbone. Furthermore, the framework can utilize background synchronization of CID updates following the initial device login, thereby eliminating potential authentication delays in subsequent login attempts.

\section{Challenges and Future Research Directions}
\label{sec:Challenges and Future Research Directions}

Our framework demonstrates superiority over existing RBA systems; however, some limitations remain. First, fairness in FL should be considered to ensure that the global model does not become biased toward the data distribution of certain contributors. Mechanisms such as contribution frequency can be employed to address this challenge \cite{ju2024accelerating}. Second, when new users join the federated setting, mechanisms like reputation scoring should be implemented to mitigate attacks such as Byzantine and Sybil attacks, where malicious entities could introduce multiple fake clients \cite{sharma2023flair}. Third, although local model weights are based on similarity scores rather than actual data values, sophisticated attackers might still infer some user data \cite{mothukuri2021survey}. Thus, privacy-preserving solutions, such as differential privacy or homomorphic encryption, are necessary to reduce the risk of data inference \cite{yang2024efficient}. Additionally, while user devices handle risk assessment and model training, server-based aggregation remains. Decentralization of this process could improve availability and eliminate single point of failure \cite{beltran2023decentralized}. Finally, although our autoencoder model performs efficiently in the federated setting for anomaly detection, investigating lightweight models or more advanced alternatives could further enhance performance \cite{saez2023clustered, ieracitano2020novel}.


\section{Conclusion}
\label{sec:Conclusion}

The F-RBA framework presents a novel approach to addressing the privacy, security, and scalability challenges inherent in conventional RBA systems. By leveraging FL, the framework enhances privacy by ensuring user data remains local, while dynamic risk assessments provide a more secure and adaptive authentication process. The use of similarity-based feature engineering further mitigates the challenges of heterogeneity in the data, enabling more accurate and consistent risk assessments across distributed devices.
Experimental results demonstrate F-RBA's superior performance over conventional unsupervised anomaly detection models, achieving a high True Positive Rate in detecting suspicious logins. It provided distributed solutions for growing user bases, showcasing adaptability to diverse contexts. While F-RBA presents advancements, future work will explore more advanced aggregation techniques and privacy-enhancing mechanisms to address remaining challenges such as fairness in FL and defenses against sophisticated attacks. The framework opens new avenues for research and offers an adaptive solution to current RBA challenges.


\bibliographystyle{IEEEtran}
\bibliography{Bibliography}


\vfill

\end{document}